\documentclass{elsart}

\usepackage{epsfig}

\usepackage{amssymb}

\begin{document}

\begin{frontmatter}



\title{Superstatistics, escort distributions, and
applications}


\author{Christian Beck}
\ead{c.beck@qmul.ac.uk}
\address{School of Mathematical Sciences \\
Queen Mary, University of London  \\
  Mile End Road, London E1 4NS, England}



\begin{abstract}
Superpositions of different statistics on different time or
spatial scales (in short, superstatistics) can naturally lead to
an effective description by nonextensive statistical mechanics. We
first discuss the role of escort distributions within the
superstatistical framework, and then briefly describe recent
physical applications of this concept to turbulent and patttern
forming systems.
\end{abstract}

\begin{keyword}
nonextensive statistical mechanics, superstatistics, escort
distributions

\PACS  05.40.-a \sep  05.70.-a

\end{keyword}

\end{frontmatter}

\section{Introduction}

In this paper we will briefly review the so-called superstatistics
concept\cite{eddie}. It yields a simple and plausible argument why
nonextensive methods, as developed by Tsallis and many others
\cite{tsallis,mendes,abe}, are relevant for nonequilibrium systems
with large-scale spatio-temporal fluctuations of an intensive
parameter.

Superstatistics is a superposition of two (or even more) different
statistics: One given by ordinary Boltzmann factors, and another
one given by large-scale fluctuations of one (or several)
intensive parameters (e.g.\ the inverse temperature). The
corresponding stationary probability distributions arise as a
convolution of the various statistics. Tsallis statistics is a
special case that arises naturally in this approach for a
$\chi^2$-distributed parameter, if the marginal distributions are
formed. But other generalized statistics are possible as well.
Recently it has become clear that the concept is not only a
theoretical construct but of practical physical relevance. For
example, it is possible to develop superstatistical models that
well describe the statistical properties of measured accelerations
in Lagrangian turbulence experiments \cite{Reynoldsprl,euro}, or
the statistics of defect velocities in pattern forming systems
\cite{daniels}, or the measured energy distribution of cosmic rays
\cite{borges,cosmic}.

\section{Basic idea of superstatistics}

Consider a driven nonequilibrium systems with spatio-temporal
fluctuations of an intensive parameter $\beta$. These fluctuations
are externally produced, by constantly putting energy into the
system which is dissipated. The intensive parameter may be the
inverse temperature, or an effective chemical potential, or a
function of the fluctuating energy dissipation in the flow (for
the turbulence application). Locally, i.e.\ in spatial regions
(cells) where $\beta$ is approximately constant, the system may be
described by ordinary statistical mechanics, i.e.\ ordinary
Boltzmann factors $e^{-\beta E}$, where $E$ is an effective energy
in each cell. In the long-term run, the system is described by a
spatio-temporal average over the fluctuating $\beta$.

Suppose the probability density of $\beta$ is $f(\beta)$. By
integrating over $\beta$ one obtains marginal distributions of the
form
\begin{equation}
p(E)=\int_0^\infty f(\beta ) \frac{1}{Z(\beta)} \rho (E)e^{-\beta
E}d\beta , \label{margi}
\end{equation}
where $\rho(E)$ is the density of states and $Z(\beta)$ is the
normalization constant of $\rho(E)e^{-\beta E}$ for a given
$\beta$. Eq.~(\ref{margi}) describes the long-term behaviour of
the system (we have chosen type-B superstatistics \cite{eddie}).

Let us consider three important examples (many more are possible).

1. $\beta$ is distributed according to the 
$\chi^2$-distribution
\begin{equation}
f (\beta) = \frac{1}{\Gamma \left( \frac{n}{2} \right)} \left\{
\frac{n}{2\beta_0}\right\}^{\frac{n}{2}} \beta^{\frac{n}{2}-1}
\exp\left\{-\frac{n\beta}{2\beta_0} \right\} . \label{fluc}
\end{equation}
$n$ is a parameter. The average of the fluctuating $\beta$ is
given by
\begin{equation}
\langle \beta \rangle =\int_0^\infty\beta f(\beta) d\beta= \beta_0
\end{equation}
and the second moment of $\beta$ is given by
\begin{equation}
\frac{\langle \beta^2 \rangle}{\beta_0^2}= 1+\frac{2}{n} .
\end{equation}
Let  $E=\frac{1}{2}u^2$ be a kinetic energy. The marginal distribution (\ref{margi}),
obtained by integration over all $\beta$, yields the generalized
canonical distributions of nonextensive statistical mechanics
\cite{tsallis,mendes,abe}
\begin{equation}
p(u) =\frac{1}{Z_q} \frac{1}{\left(
1+\frac{1}{2}\tilde{\beta}(q-1)u^2\right)^{\frac{1}{q-1}}}
\label{gencan}
\end{equation}
where $Z_q$ is a normalization constant and
\begin{eqnarray}
q&=&1+\frac{2}{n+1}\label{qnn}
\\
\tilde{\beta}&=&\frac{2}{3-q} \beta_0.
\end{eqnarray}
The distributions (\ref{gencan}) maximize the Tsallis entropies
$S_q$ subject to suitable constraints \cite{tsallis,mendes,abe}.

2. $\beta$ is log-normally distributed:
\begin{equation}
f(\beta) = \frac{1}{\beta s \sqrt{2\pi}}\exp\left\{ \frac{-(\log
\frac{\beta}{m})^2}{2s^2}\right\}
\end{equation}
The average $\beta_0$ of the above log-normal distribution is
given by $\beta_0=m\sqrt{w}$ and the variance by
$\sigma^2=m^2w(w-1)$, where $w:= e^{s^2}$. One obtains the
superstatistics distribution
\begin{equation}
p(u) = \frac{1}{2\pi s }\int_0^\infty d\beta \; \beta^{-1/2}
\exp\left\{ \frac{-(\log \frac{\beta}{m})^2}{2s^2}\right\}
e^{-\frac{1}{2}\beta u^2}.  \label{10}
\end{equation}
The integral cannot be evaluated in closed form, but it can be
easily numerically evaluated and compared with experimental data
\cite{Reynoldsprl,euro,castaing}.

3. Yet another interesting example of a superstatistics is
obtained if we assume that the temperature $T=\beta^{-1}$ (rather
than $\beta$ itself) is $\chi^2$-distributed. In this case the
corresponding marginal distribution $p(u)$ has exponential tails
\cite{sattin}.

\section{Escort distributions in superstatistical systems}

For a given probability distribution $\{p_i\}$ of microstates $i$
and a given parameter $q$ the escort distribution $\{P_i\}$ is defined
as \cite{BS}
\begin{equation}
P_i=\frac{p_i^q}{\sum_i p_i^q}.
\end{equation}
Both types of distributions, $\{p_i\}$ and $\{P_i\}$, notoriously
occur in the formulation of nonextensive statistical mechanics
\cite{mendes,abe,abeescort}, like two spin-degrees of freedom
which are not there in classical statistical mechanics ($q=1$). We
will now show that within the superstatistical approach these two
degrees of freedom can be easily understood.

Generally, it makes sense to $\beta$-average typical thermodynamic
relations if the local cells are sufficiently large and if they
are in local equilibrium. In this case one has for each local cell
\begin{equation}
S=\beta (U-F),
\end{equation}
where $S$ is the ordinary Boltzmann-Gibbs-Shannon entropy, $U$ is
the internal energy and $F$ the free energy associated with each
cell. The averaged version reads
\begin{equation}
\int_0^\infty f(\beta) S d\beta =\int_0^\infty \beta f(\beta)U
d\beta -\int_0^\infty \beta f(\beta) F d\beta . \label{gibbs}
\end{equation}
Note that on the right-hand side we obtain averages formed with
$\beta f(\beta)$. For $\chi^2$-distributed $\beta$ this is like
transforming $n\to n+2$ in the distribution (\ref{fluc}). Since
\begin{equation}
\frac{1}{q-1}=\frac{n+1}{2}
\end{equation}
and
\begin{equation}
\frac{q}{q-1}=\frac{n+3}{2}
\end{equation}
this means one is forming averages with respect to the escort
distributions on the right-hand side of eq.~(\ref{gibbs}), as
required in the modern versions of nonextensive statistical
mechanics \cite{mendes}.

Our argument is valid for arbitrary energies $E$. We see that
within the superstatistical $\beta$-averaging
approach entropic quantities most naturally correspond to
the original distribution $p_i$, energetic
quantities such as $U$ or $F$ most naturally to the escort
distribution $P_i$, since there is an additional factor $\beta$ on the
right-hand side of eq.~(\ref{gibbs}).

In general, the averaged Shannon entropy on the left-hand side
does not coincide with the Tsallis entropy $S_q$, which satisfies
\cite{tsallis,mendes,abe}
\begin{equation}
S_q=\tilde{\beta} (U_q-F_q). \label{gibbsq}
\end{equation}
However, it might be interesting to construct special situations
where eq.~(\ref{gibbs}) and (\ref{gibbsq}) are equivalent.

\section{Applications}

Langevin equations where the parameters fluctuate on a large
spatio-temporal scale yield a concrete dynamical realization of
superstatistics \cite{prl}. For example, to model the motion of a
single test particle in a turbulent flow one considers a
superstatistical extension of the Sawford model \cite{saw}. The
Sawford model describes the joint stochastic process
$(a(t),u(t),x(t))$ of acceleration, velocity and position of a
Lagrangian test particle in a turbulent flow by the stochastic
differential equation
\begin{eqnarray}
\dot{a}& =&-(T_L^{-1}+t_\eta^{-1})a-T_L^{-1}t_\eta^{-1} u
\nonumber \\ &\,&
+\sqrt{2\sigma_u^2(T_L^{-1}+t_\eta^{-1})T_L^{-1}t_\eta^{-1}}\;
L(t)
\\ \dot{u} &=&a \\ \dot{x} &=&u,
\end{eqnarray}
where $L(t)$ is Gaussian white noise. Note that in this model the
acceleration $a$ and not the velocity $u$ is driven by white
noise, so the meaning of the variables is different as compared to
an ordinary Brownian particle. $T_L$ and $t_{\eta}$ are two time
scales, with $T_L
>>t_\eta$ and
\begin{eqnarray}
T_L&=& \frac{2\sigma_u^2}{C_0 \epsilon} \\ t_\eta &=&
\frac{2a_0\nu^{1/2}}{C_0\epsilon^{1/2}}. \\
\end{eqnarray}
$\epsilon$ is the energy dissipation, $\nu$ the kinematic
viscosity, $C_0, a_0$ are Lagrangian structure function constants,
and $\sigma_u^2$ is the variance of the velocity distribution. The
Sawford model with constant coefficients predicts Gaussian
stationary distributions for $a$ and $u$, and is thus at variance
with recent measurements \cite{boden}. However, a superstatistical
generalization of the Sawford model \cite{Reynoldsprl,euro} well
fits the data.


For $T_L \to \infty$ one can derive \cite{euro} that the
fluctuating parameter $\beta$ is given by
\begin{equation}
\beta = \frac{2a_0}{C_0^2} \nu^{1/2} \epsilon^{-3/2}, \label{26}
\end{equation}
i.e. the fluctuating energy dissipation $\epsilon$ is proportional
to $\beta^{-2/3}$. Fig.~1 shows the measured probability density
of the acceleration of a Lagrangian test particle in a turbulent
flow as obtained in the experiment of Bodenschatz et al.
\cite{boden}. Log-normal superstatistics with $s^2=3.0$ yields an
excellent fit \cite{euro}.

\begin{figure}

\epsfig{file=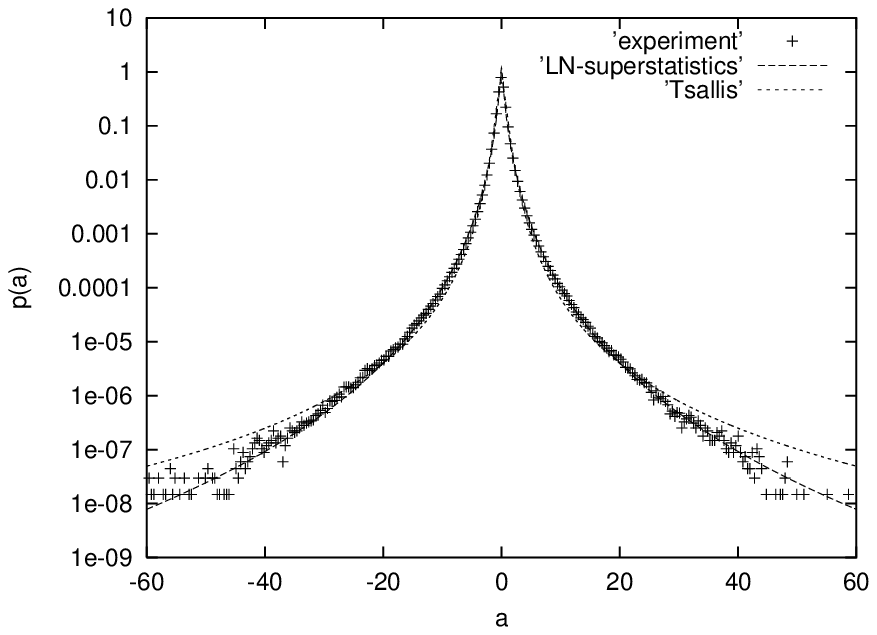}

{\bf Fig.~1} Histogram of accelerations $a$ as measured in
\cite{boden} and the log-normal superstatistics prediction
eq.~(\ref{10}) with $s^2=3.0$. For comparison, the figure also
shows Tsallis statistics (eq.~(\ref{gencan}) with $q=1.5$).

\end{figure}

Similar superstatistical models can be formulated for pattern
forming systems, just that the meaning of the variables in the
Langevin equation is different. The probability density of defect
velocities in inclined layer convection experiments can be quite
precisely measured. As shown in Fig.~2, it quite precisely
coincides with a Tsallis distribution with $q \approx 1.46$
\cite{daniels}. A superstatistical model well describes 
various stochastic properties of these defects.

Superstatistical models also have applications in high energy
physics, for example as simple models to explain the measured
statistics of cosmic rays \cite{borges,cosmic}.
\begin{figure}

\epsfig{file=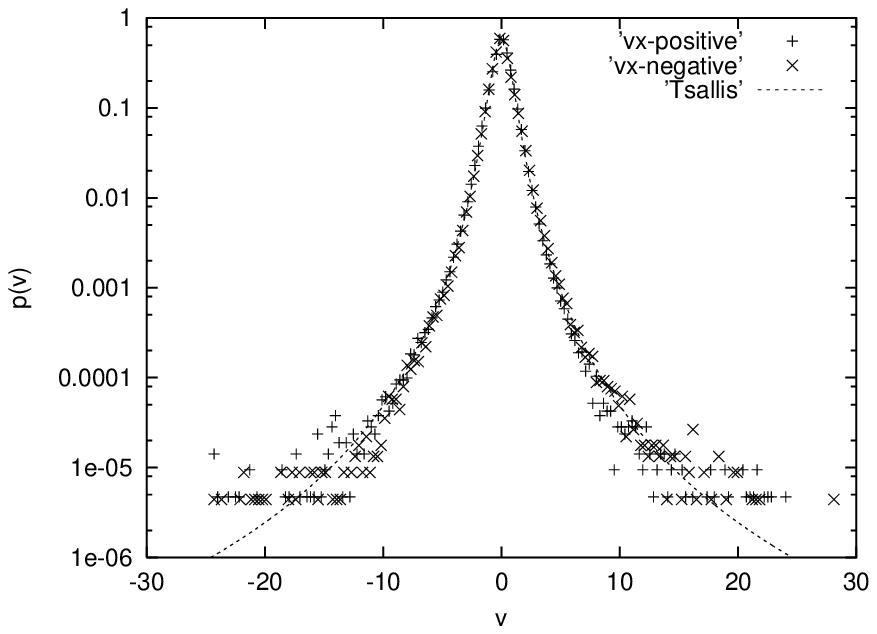}

{\bf Fig.~2} Measured distribution of positive and negative defect
velocities in inclined layer convection and comparison with a
nonextensive canonical distribution of type (\ref{gencan}) with
$q=1.46$ (more details in \cite{daniels}).

\end{figure}


\begin{thebibliography}{99}
\bibitem{eddie} C. Beck and E.G.D. Cohen, Physica {\bf 322A},
267 (2003)
\bibitem{tsallis} C. Tsallis, J. Stat. Phys. {\bf 52}, 479 (1988)
\bibitem{mendes} C. Tsallis,  R.S. Mendes and A.R. Plastino, Physica {\bf 261A},
534 (1998)
\bibitem{abe} S. Abe, Y. Okamoto (eds.),
{\it Nonextensive Statistical Mechanics and Its Applications},
Springer, Berlin (2001)
\bibitem{Reynoldsprl} A.M. Reynolds, Phys. Rev. Lett. {\bf 91}, 084503 (2003)
\bibitem{euro} C. Beck, Europhys. Lett. {\bf 64}, 151 (2003)
\bibitem{daniels} K.E. Daniels, C. Beck, and E. Bodenschatz,
Physica D, in press (2003) (cond-mat/0302623)
\bibitem{borges} C. Tsallis, J.C. Anjos, E.P. Borges,
Phys. Lett. {\bf 310A}, 372 (2003) 
\bibitem{cosmic} C. Beck, Physica {\bf 331A}, 173 (2004)
\bibitem{castaing} B. Castaing, Y Gagne, E.J. Hopfinger,
Physica {\bf 46D},
177 (1990)
\bibitem{sattin} F. Sattin and L. Salasnich,
Phys. Rev. {\bf 65E}, 035106(R) (2002)
\bibitem{BS} C. Beck and F. Schl\"{o}gl, {\em
Thermodynamics of Chaotic Systems}, Cambridge University Press,
Cambridge (1993)
\bibitem{abeescort} S. Abe, Phys. Rev. {\bf 68E}, 031101 (2003)


\bibitem{prl} C. Beck, Phys. Rev. Lett. {\bf 87}, 180601 (2001)



\bibitem{saw} B.L. Sawford, Phys. Fluids {\bf A3}, 1577 (1991)


\bibitem{boden} G.A. Voth et al., J. Fluid Mech. {\bf 469}, 121 (2002)






\end{thebibliography}
\end{document}